\documentclass[10pt,conference]{IEEEtran}
\IEEEoverridecommandlockouts
\usepackage{cite}
\usepackage{amsmath,amssymb,amsfonts}
\usepackage{algorithmic}
\usepackage{graphicx}
\usepackage{textcomp}
\usepackage{xcolor}
\usepackage{cite}
\usepackage{amsmath,amssymb,amsfonts}
\usepackage{algorithmic}
\usepackage{graphicx}
\usepackage{textcomp}
\usepackage{subcaption}
\usepackage{hyperref}
\usepackage{bm}
\usepackage{amsmath}
\usepackage{multirow}
\usepackage{booktabs}
\usepackage{multirow}
\usepackage{booktabs}
\usepackage{colortbl}
\usepackage{xcolor}
\usepackage{changepage} 
\usepackage[table]{xcolor}

\def\BibTeX{{\rm B\kern-.05em{\sc i\kern-.025em b}\kern-.08em
    T\kern-.1667em\lower.7ex\hbox{E}\kern-.125emX}}
\begin{document}

\title{Chart2Code-MoLA: Efficient Multi-Modal Code Generation via Adaptive Expert Routing}

\author{
\IEEEauthorblockN{
   Yifei Wang, Jacky Keung, Zhenyu Mao, Jingyu Zhang*, Yuchen Cao
}
\IEEEauthorblockA{Department of Computer Science, City University of Hong Kong, Hong Kong, China\\
    ywang4748-c@my.cityu.edu.hk, Jacky.Keung@cityu.edu.hk, zhenyumao2-c@my.cityu.edu.hk, \\ 
    jzhang2297-c@my.cityu.edu.hk, cynthcao2-c@my.cityu.edu.hk
}
\thanks{* Corresponding author: Jingyu Zhang.}
}
\maketitle

\begin{abstract}
Chart-to-code generation is a critical task in automated data visualization, translating complex chart structures into executable programs. 
While recent Multi-modal Large Language Models (MLLMs) improve chart representation, existing approaches still struggle to achieve cross-type generalization, memory efficiency, and modular design. 
To address these challenges, this paper proposes C2C-MoLA, a multimodal framework that synergizes Mixture of Experts (MoE) with Low-Rank Adaptation (LoRA). 
The MoE component uses a complexity-aware routing mechanism with domain-specialized experts and load-balanced sparse gating, dynamically allocating inputs based on learnable structural metrics like element count and chart complexity. 
LoRA enables parameter-efficient updates for resource-conscious tuning, further supported by a tailored training strategy that aligns routing stability with semantic accuracy. 
Experiments on Chart2Code-160k show that the proposed model improves generation accuracy by up to 17\%, reduces peak GPU memory by 18\%, and accelerates convergence by 20\%, when compared to standard fine-tuning and LoRA-only baselines, particularly on complex charts.
Ablation studies validate optimal designs, such as 8 experts and rank-8 LoRA, and confirm scalability for real-world multimodal code generation.
\end{abstract}


\section{Introduction}
With the growing reliance on data visualization for insight communication, automated understanding of visual content has become central to intelligent systems\cite{wu2023ai}.
Chart-to-code generation refers to the task of converting visualizations such as bar, line, and scatter plots into executable programs, which requires precise interpretation of visual semantics and syntactically correct code generation \cite{liu2023mmc}.
Early methods used rule-based extraction from synthetic images or well-structured images\cite{minu2014semantic}\cite{gevaert2018deep}, leveraging predefined heuristics to identify chart elements.
Subsequently, neural approaches introduced encoder-decoder architectures to tackle chart-to-data tasks\cite{sviatov2021data}\cite{liang2024automatic}, learning to map pixel-level inputs to structured representations. These advances laid the foundation for multimodal code generation by bridging visual and symbolic modalities\cite{chakraborty2021multi}.

Recent progress has been driven by Multimodal Large Language Models (MLLMs) that leverage visual-language pretraining and instruction tuning\cite{han2023chartllama}\cite{zhang2025enhancing}. 
These models have demonstrated notable effectiveness in chart understanding tasks such as ChartQA \cite{xu2025chartmoe} and chart-to-code translation \cite{zhao2025chartcoder}, benefiting from scalable and transferable training paradigms. 
Nevertheless, current methods face key limitations: 1) poor generalization to complex charts such as grouped bar or multi-series plots \cite{han2023chartllama};
2) high GPU memory requirements due to full fine-tuning, often exceeding 15GB \cite{li2024mmcode}, as all parameters require storing gradients and optimizer states;
3) monolithic model designs that lack modularity, limiting both interpretability and adaptability \cite{he2024llms}. These issues necessitate a balanced approach to specialization and efficiency for structurally diverse inputs.

To address these challenges, this paper proposes Chart-to-Code Mixture of Low-rank Adaptation (C2C-MoLA), a multimodal framework that integrates Mixture of Experts (MoE) with Low-Rank Adaptation (LoRA) for chart-to-code generation. 
The framework is designed to handle structurally diverse charts  by assigning inputs adaptively to specialized expert modules and fine-tuning large models efficiently under constrained computational budgets.
To ensure stable and scalable training, we introduce a tailored optimization strategy for modular architectures. 
C2C-MoLA leverages multimodal inputs  including  visual chart images and chart-type metadata to generate syntactically correct and semantically faithful code, improving generalization, memory efficiency, and modularity over existing approaches.

Specifically, C2C-MoLA uses a structure-aware routing mechanism to assign inputs to eight domain-specialized experts via sparse gating, guided by a learnable complexity metric based on chart type and visual element density. 
To promote balanced expert utilization and prevent routing collapse, it injects Gaussian noise into gating scores and applies an auxiliary load-balancing constraint.
LoRA enables efficient adaptation by applying low-rank updates to attention layers. Training further integrates syntax-aware and semantic reconstruction losses, KL-based expert regularization, and memory-efficient scheduling to support stable optimization of the architecture. Together, these components enable accurate and efficient chart-to-code generation. 
The main contributions of this paper include:
\begin{itemize}
    \item Proposed C2C-MoLA, a multimodal chart-to-code generation framework that integrates MoE and LoRA to address challenges in generalization, efficiency, and modularity.
    \item Designed a complexity-aware routing strategy based on a learnable structural metric, enabling adaptive expert assignment for diverse chart types.
    \item Developed a training strategy that combines syntax- and semantics-aware objectives, expert regularization, and memory-efficient scheduling to support scalable optimization.
    \item Demonstrated consistent performance improvements over strong baselines, with ablation studies validating the contribution of each component.
\end{itemize}

\section{Background and Related Work}
\subsection{Chart Understanding and Analysis}
Chart understanding is fundamental to automated visual data interpretation, particularly in chart-to-code generation systems. 
The process begins with chart type recognition, which involves identifying categories such as bar, line, pie, and scatter charts, along with their structural layouts\cite{huang2024pixels}. This recognition serves as the semantic anchor for downstream tasks\cite{farahani2023automatic}. 
Traditional methods used Convolutional neural networks to extract spatial features for classification\cite{zhai2022scaling}, but they often struggled with long-range dependencies and complex element relationships, such as overlapping legends in grouped bar charts\cite{liu2021swin}. 
These limitations led to the adoption of transformer-based models, where self-attention mechanisms dynamically focus on relevant visual regions and significantly improve recognition for intricate layouts\cite{khan2022transformers}.

Beyond classification, precise element extraction is critical. Charts consist of components such as axes, legends, titles, labels, and data points, which define the semantic structure\cite{davila2020chart}.
Early techniques relied on image processing methods such as edge detection \cite{yu2018simultaneous} and segmentation\cite{jung2017chartsense}, but these often struggled with ambiguous or overlapping elements.
Modern deep learning approaches that combine convolutional networks for local feature extraction with transformers for contextual reasoning have shown superior performance in detection and localization accuracy\cite{lu2024separable}. 
These methods produce structured inputs for code generation, enabling faithful reconstruction using libraries such as Matplotlib\cite{hunter2007matplotlib} or Plotly\cite{sievert2020interactive}.

As visualizations grow more complex, multi-modal chart understanding becomes increasingly important. 
Most charts embed textual elements such as axis labels and annotations within visual structures, requiring modeling of layout and semantics\cite{xiao2023let}.
Vision-language architectures, including dual-attention transformers, align spatial distributions with textual cues\cite{wang2023image}.
For example, when interpreting a bar chart, a model must align spatial bar distributions with corresponding text labels to produce accurate executable code.

Ultimately, effective chart understanding requires unified modeling of both numerical distributions and textual annotations, as charts often embed semantic labels directly within visual layouts\cite{huang2024pixels}. 
This demands not only accurate visual recognition but also cross-modal reasoning to align visual elements with their corresponding text.
Recent advances further enhance this paradigm by introducing component-specific attention mechanisms that explicitly focus on key chart elements such as axes, labels, and legends \cite{hou2025multimodal}.
These advances form a solid foundation for modular and multi-modal systems capable of handling the complexity of chart-to-code generation.

\subsection{Vision-to-Code Generation}
Automated code generation from visual inputs has emerged as a pivotal machine learning domain, driven by demands for automating web interface development and data visualization workflows\cite{de2020recent}. 
This field spans diverse applications including GUI to HTML/CSS conversion\cite{zhekova2023automatic}, flowchart to program translation\cite{liu2022code}, and crucially chart to executable code generation. 
These tasks require models to jointly interpret visual layouts and semantic functions of elements, ensuring that the generated code accurately reflects the intended behavior.

Within chart-to-code generation, methodologies have progressed through several phases, including template-based methods, machine learning approaches, and multimodal fusion techniques. 
Early template-based methods matched predefined code snippets to specific chart structures \cite{minu2014semantic}; while efficient for simple charts, they lacked flexibility for novel designs. 
Subsequently, machine-learning methods leveraged CNNs and transformers to recognize chart elements such as axes and legends, mapping them to code constructs\cite{sviatov2021data}; although more capable, these models often struggle with intricate data relationships and dynamic interactivity\cite{liang2024automatic}. 
Most recently, multimodal fusion techniques have sought to align visual and textual elements, yet still face generalization challenges across diverse chart types, due to limited handling of complex layouts, weak visual-semantic alignment, high fine-tuning costs, and poor adaptability to novel patterns\cite{he2024llms}. 

As a prominent representative of recent multimodal approaches, large language models (LLMs) have significantly advanced code generation capabilities. Systems such as GPT-4.5 and Claude 3 Opus support long-context reasoning and multi-turn synthesis, while open-source alternatives like Code LLaMA~\cite{touvron2023llama}, DeepSeek-Coder~\cite{guo2024deepseek}, and Qwen-Code~\cite{bai2023qwen} offer greater flexibility through fine-tuning and local deployment. 
Despite these gains, current LLMs still struggle with domain-specific generalization, particularly in tasks involving structured or multimodal inputs.
AgentCoder\cite{huang2023agentcoder}, through multi-agent collaboration, and WizardCoder\cite{luo2023wizardcoder}, via instruction tuning, show improved alignment with user intent, yet remain incapable of handling layout-sensitive or visually grounded chart generation tasks.

Recent research addresses these through modular architectures enabling adaptive reasoning. Particularly promising is the Mixture of Experts (MoE) framework, which achieves specialization while maintaining efficiency through sparse activation\cite{mu2025comprehensive}. This naturally extends to multimodal code generation where distinct experts can handle varied visual textual patterns.

\subsection{Mixture-of-Experts and Routing}
The Mixture of Experts architecture represents a modular deep learning paradigm that achieves scalable modeling through dynamic input routing to specialized subnetworks\cite{fedus2022review}. 
To address the structural diversity and semantic complexity of chart inputs, MoE has gained increasing attention for its scalability and adaptability. 
Unlike conventional models that activate all parameters for every input, MoE selectively routes each instance to a small subset of experts using a gating function\cite{mu2025comprehensive}. This sparse activation preserves model capacity while reducing computational cost, making it well-suited for heterogeneous tasks like chart-to-code generation, where different chart types demand tailored processing strategies.

Each expert in an MoE model is trained to specialize in distinct input characteristics, such as spatial layouts or annotation styles\cite{zhong2022meta}, enabling more accurate pattern modeling and faster convergence. However, expert overuse and underutilization often lead to load imbalance, which degrades model capacity and convergence stability\cite{zeng2025efficientmoe}. Auxiliary routing losses and load-balancing regularization are commonly used to address this\cite{liu2023sparse}, encouraging equitable expert utilization and improving overall training efficiency.

Despite these architectural strengths, the effectiveness of MoE hinges critically on the design of routing mechanisms. Prior approaches often rely on fixed or hand-crafted rules. For instance, ChartMoE statically assigns experts based on chart type\cite{xu2025chartmoe}, providing a coarse-grained partition that overlooks intra-class variation. ChartCoder\cite{zhao2025chartcoder} employs multimodal routing but depends on hardcoded visual features, lacking adaptability to unseen structural patterns. The Dual-Preference model\cite{zhang2025enhancing} optimizes generation through iterative tuning, yet disregards input structure in routing. Route-to-Reason (RTR)\cite{pan2025route} introduces dynamic expert selection under computation budgets, but makes routing decisions based on latent token-level signals rather than interpretable chart structures.

A summary of these methods is shown in Tab.~ \ref{tab:method_comparison}, comparing their routing design, structural awareness, and integration of MoE with parameter-efficient adaptation. In contrast, our proposed C2C-MoLA introduces a dynamic routing module that leverages interpretable structural cues for fine-grained expert selection and enables seamless synergy between LoRA-based adaptation and MoE specialization.
\begin{table}[t]
\centering
\caption{Chart-to-Code Model Comparison}
\begin{tabular}{l|c|c|c}
\toprule
\multirow{2}{*}{Method} & Routing & Structural & LoRA-MoE \\
& Awareness & Metric & Synergy \\
\midrule
ChartCoder\cite{zhao2025chartcoder}  & Static & $\times$ & $\times$ \\
ChartMoE\cite{xu2025chartmoe}   & Static  & $\times$ & $\times$ \\
Dual-Preference\cite{zhang2025enhancing}  & $\times$ & $\times$ & $\times$ \\
Route-to-Reason\cite{pan2025route}  & Dynamic & $\times$ & $\times$ \\
\textbf{C2C-MoLA (Ours)}         & \textbf{Dynamic} & \checkmark & \checkmark \\
\bottomrule
\end{tabular}
\label{tab:method_comparison}
\end{table}

\subsection{Parameter-Efficient Fine-Tuning}
Parameter-efficient fine-tuning has emerged as a critical strategy for adapting large pretrained models to downstream tasks while minimizing retraining costs\cite{ding2023parameter}. 
Among various methods, Low-Rank Adaptation (LoRA) has demonstrated particular effectiveness through low rank matrix decomposition of weight updates.
Rather than modifying the entire parameter space, LoRA introduces lightweight trainable matrices $\Delta W = AB$, with $B \in \mathbb{R}^{r \times d}$, $A \in \mathbb{R}^{d \times r}$, and $r \ll d$, capturing task-specific knowledge through minimal updates\cite{lin2024tracking}.
This approach drastically reduces computation and memory demands, making it ideal for multimodal tasks like chart-to-code generation where simultaneous visual-textual processing requires efficient resource utilization\cite{wang2025parameter}.  

When compared to alternative methods, LoRA provides a superior balance between efficiency and expressiveness. Prefix tuning optimizes soft prompts at the input level but may underfit complex internal representations\cite{li2021prefix}.
Adapters inject additional neural modules between layers, increasing flexibility at the cost of expanded model size and training time\cite{he2021effectiveness}. 
BitFit updates only bias terms to achieve maximal efficiency but struggles to capture nuanced patterns\cite{xu2023parameter}. 
In contrast, LoRA directly modifies the weight space through low-rank projections, enabling efficient adaptation without sacrificing model capacity\cite{zhang2023lora}. This characteristic makes it particularly advantageous for specialized applications demanding both precision and scalability.  

LoRA offers not only parameter efficiency but also improved memory usage and faster convergence, enabling large models to be fine-tuned under limited GPU resources. This makes it especially suitable for lightweight and scalable real-time applications. When combined with LoRA, Mixture-of-Experts (MoE) models can retain adaptability even under constrained computational budgets\cite{zhang2024mpmoe}. Recent developments such as LLaMA-Adapter showcase LoRA’s effectiveness in vision-language tasks\cite{zhang2023llama}, while MoRA\cite{tang2024mor} and Little by Little\cite{lu2025little} attempt to integrate sparse activation with efficient tuning. However, these approaches often rely on static expert assignment or manually designed adapter placement, limiting their flexibility in handling complex or diverse input structures.

These limitations motivate C2C-MoLA’s design, which integrates dynamic expert routing with LoRA adaptation, thereby enabling input-aware specialization while maintaining computational scalability. This design philosophy aims to bridge the gap between general-purpose multimodal large models and the specialized demands of chart code generation.

\section{proposal}
\label{sec:proposal}
To address three core challenges including poor generalization, high GPU usage, and monolithic design, we propose C2C-MoLA, a multi-stage framework that integrates adaptive expert routing and parameter-efficient adaptation. 
The MoE-based router enables input-aware specialization across diverse chart types, improving generalization and modularity, while LoRA reduces memory overhead during fine-tuning.

\begin{figure*}[hbtp]
    \centering
    \includegraphics[width=\textwidth]{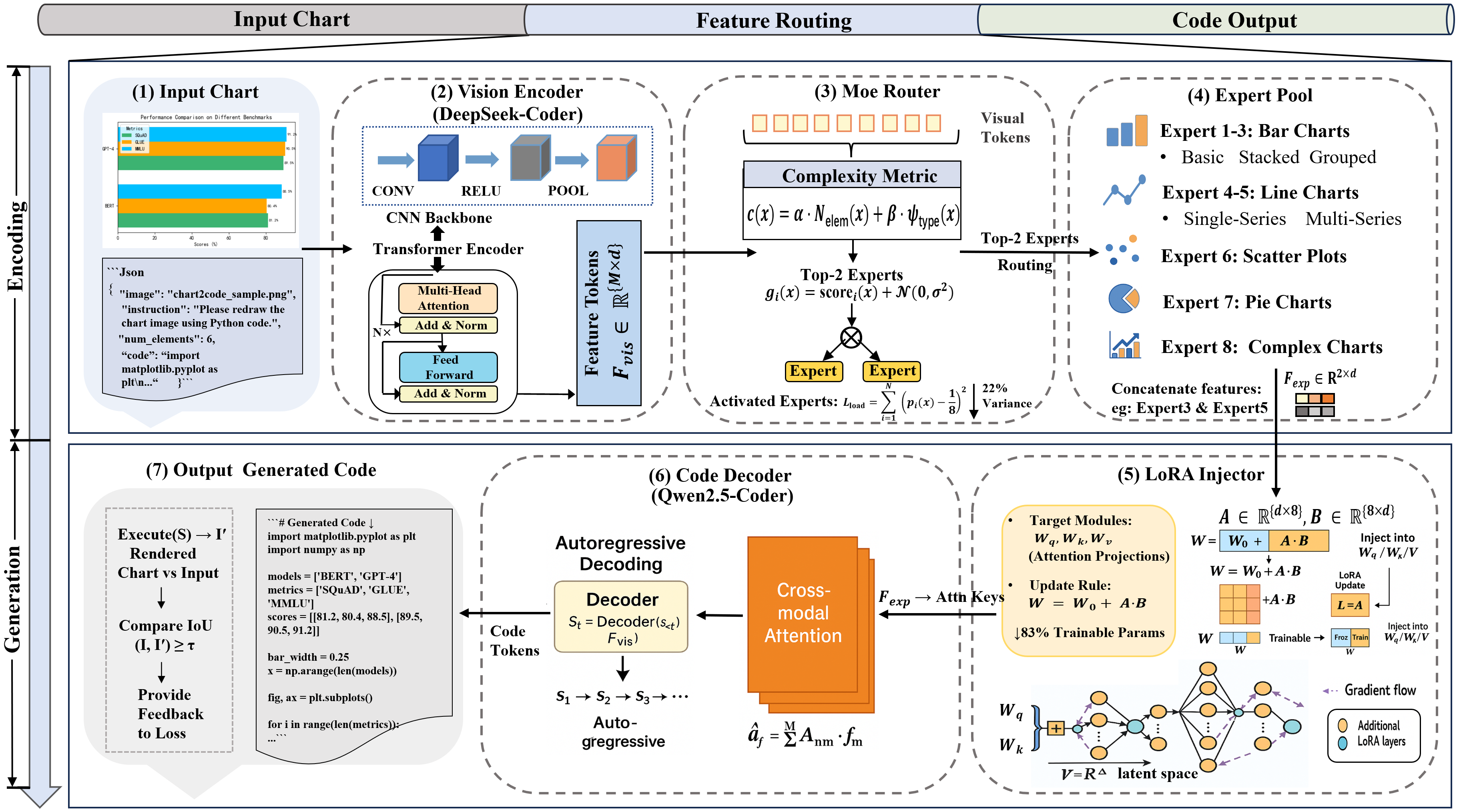}
    \caption{C2C-MoLA: Chart-to-Code Generation Framework}
    \label{fig:1}
\end{figure*}

Fig.~\ref{fig:1} illustrates the pipeline, which starts with a chart image and metadata such as chart type and element count. 
This input is encoded by a dual-stream vision backbone (\textit{DeepSeek-Coder}) combining convolutional and transformer layers to extract hierarchical visual tokens $F_{\text{vis}}$. 
These tokens are routed through a complexity-aware MoE module, where routing is guided by a learned structural complexity metric.
The top-2 specialized experts are selected based on this score.
Their outputs are then fused and injected with LoRA-based low-rank updates (applied to attention projections $W_q$, $W_k$, $W_v$), and passed to an autoregressive decoder (\textit{Qwen2.5-Coder}). 
A cross-modal attention mechanism aligns the visual features with the generated code tokens to produce executable Python code. Finally, semantic fidelity is validated by rendering the output and measuring visual similarity with the input chart validated using an IoU threshold, where $\text{IoU}(I, I') \geq \tau$.

To summarize, this formulation directly addresses three key challenges in chart-to-code generation.  
First, it improves modality alignment between the heterogeneous input $x$ and output sequence $S$ by enforcing the semantic consistency constraint (see Sec.~III-A).  
Second, it supports scalable learning via complexity-aware expert routing and LoRA-based parameter adaptation (see Sec.~III-B and III-C).  
Third, it balances semantic fidelity and syntactic correctness through the joint optimization of semantic rendering accuracy and a token-level syntax loss (see Sec.~III-D).

\subsection{Problem Definition}
The chart-to-code generation task is formally defined as learning a mapping:
\begin{equation}
f: \mathbb{R}^{H \times W \times C} \rightarrow \mathcal{Y}
\label{eq1}
\end{equation}

where the input is a chart image $I \in \mathbb{R}^{H \times W \times C}$, and the output is a token sequence $S = [s_1, s_2, \dots, s_n] \in \mathcal{Y}$, representing executable Python code. 

The output must be both syntactically correct (i.e., valid Python code), and semantically faithful. That is, the visualization rendered by executing $S$ must align with the original chart image $I$.

Semantic fidelity is formalized through a validity constraint:
\begin{equation}
C_{\text{semantics}} = \left\{ S \,\middle|\, \text{execute}(S) \rightarrow I' \text{ and } \text{IoU}(I, I') \geq \tau \right\}
\label{eq2}
\end{equation}

where $\text{IoU}(I, I')$ computes the Intersection-over-Union similarity between original and rendered charts, with $\tau \in [0.75, 0.90]$ as the fidelity threshold \cite{yang2024chartmimic}. The set $C_{\text{semantics}}$ represents all code sequences $S$ that produce renderings $I'$ semantically consistent with the input chart $I$.

To enable dynamic expert selection in MoE (Sec. III-B), a learnable complexity metric is defined as:
\begin{equation}
c(\mathbf{x}) = \alpha \cdot \underbrace{N_{\text{elem}}(\mathbf{x})}_{\text{element count}} + \beta \cdot \underbrace{\psi_{\text{type}}(\mathbf{x})}_{\text{type complexity}}
\label{eq:complexity}
\end{equation}

where \( \mathbf{x} = \text{VisionEncoder}(I) \) is the visual feature extracted from the input chart, \( N_{\text{elem}}(\mathbf{x}) \) counts the number of visual elements (e.g., bars, ticks, legends), and \( \psi_{\text{type}}(\mathbf{x}) \) estimates the inherent complexity of the chart type. Scalars \( \alpha \) and \( \beta \) are learnable weights that adaptively balance the two components. This metric serves as the basis for dynamic expert selection, detailed in Sec.~III-B.

\subsection{ MoE-Enhanced Architecture Design}
As shown in Fig.~\ref{fig:1}, the Mixture of Experts (MoE) module in C2C-MoLA integrates two key components: 
a complexity-aware routing mechanism and a domain-specialized expert pool. 
The routing mechanism assigns visual tokens to the top-2 experts, selected from the expert pool based on a learned complexity metric $c(x)$, which considers both the number of visual elements and chart type difficulty. 
As summarized in Tab.~\ref{tab:11} , expert allocation follows the chart-type distribution in the training data, with more frequent chart types (e.g., bar and line charts) receiving greater specialization. 
This ensures that the model can adapt to both common and rare chart types while maintaining computational efficiency and semantic fidelity.
\begin{table}[h]
\centering
\caption{Domain-Specialized Expert Configuration}
\label{tab:11}
\begin{tabular}{l|l|l|l}
\toprule
\textbf{Chart Type} & \textbf{Experts} & \textbf{Specialization} & \textbf{Ratio} \\
\midrule
Bar Charts     & 3 & Basic, Stacked, Grouped & 31.3\% \\
Line Charts    & 2 & Single-Series, Multi-Series & 26.8\% \\
Scatter Plots  & 1 & Correlation Visualization & 17.9\% \\
Pie Charts     & 1 & Proportional Representation & 13.4\% \\
Complex Charts & 1 & Hybrid or Atypical Layouts & 8.9\% \\
\bottomrule
\end{tabular}
\label{tab:11}
\vspace{2pt}
\caption*{\footnotesize * Ratio indicates the percentage of training samples for each chart type.}
\end{table}

\subsubsection{ Base Model Architecture}
The dual-stream design comprises a vision encoder and a code generator. The vision encoder, based on the DeepSeek-Coder backbone combining CNN and Transformer layers, transforms the input image $I \in \mathbb{R}^{H \times W \times C}$ into a sequence of visual tokens: 
\begin{equation}
F_{\text{vis}} = \text{VisionEncoder}(I) \in \mathbb{R}^{M \times d}
\label{eq4}
\end{equation}

where $M$ denotes the number of tokens and $d$ is the token dimension.

The code generator, implemented using the Qwen2.5-Coder decoder, autoregressively generates code tokens $\{s_n\}_{n=1}^{N}$. 
To bridge modalities, a cross-modal attention mechanism (see Fig.~\ref{fig:1}, black arrows) aligns the visual features with the generated code tokens. At each decoding step $n$, attention scores between the current token $s_n$ and each visual token $f_m$ are computed as:
\begin{equation}
\text{score}(s_n, f_m) = \text{Linear}\left([\text{PE}(s_n) \oplus \text{PE}(f_m)]\right)
\label{eq5}
\end{equation}

where $\text{PE}(\cdot)$ denotes positional encoding and $\oplus$ represents vector concatenation.

These scores are normalized using softmax to produce the attention weights:
\begin{equation}
A_{nm} = \frac{\exp(\text{score}(s_n, f_m))}{\sum_{m'=1}^{M} \exp(\text{score}(s_n, f_{m'}))}
\label{eq6}
\end{equation}

The attended context vector \( \hat{f}_n \) is then computed as a weighted sum of visual tokens:
\begin{equation}
\hat{f}_n = \sum_{m=1}^{M} A_{nm} \cdot f_m
\label{eq8}
\end{equation}

This mechanism enables the decoder to focus on relevant visual regions when generating each token, ensuring semantic alignment between image content and code output.

Each expert $E_i$ is a fine-tuned FFN optimized for its domain. To support adaptive routing, the expert selection process is guided by a complexity-aware reweighting mechanism. Given input features $x$, a gating network computes initial expert probabilities $p_i(x)$, which are then adjusted based on a structural complexity score $c(x)$ using temperature-scaled softmax:
\begin{equation}
p'_i(x) = \frac{p_i(x) \cdot \exp\left(c(x)/T\right)}{\sum_j p_j(x) \cdot \exp\left(c(x)/T\right)}
\label{eq:complexity_softmax}
\end{equation}

Here, \( T \) denotes a softmax temperature that controls the specialization intensity in expert routing.
A higher complexity score \( c(x) \) amplifies expert probabilities, encouraging routing toward more capable experts; conversely, a lower \( c(x) \) reduces this effect, allowing the base routing distribution to dominate.

The final output is computed as a weighted combination of expert responses:
\begin{equation}
y_{\text{final}} = \sum_{i=1}^{N} p'_i(x) \cdot y_i
\label{eq:expert_output}
\end{equation}

where $y_i = E_i(x)$ is the output of the $i$-th expert, and $p'_i(x)$ is the adjusted routing probability. This routing strategy aligns expert capacity with task complexity, improving both precision and generalization.

\subsubsection{ Routing Mechanism}
To enable adaptive expert selection while maintaining computational efficiency, a softmax-based sparse gating mechanism is employed, where only the top-$k$ experts are activated for each input. For a given input feature vector $x$, the gating score for expert $i$ is computed as:
\begin{equation}
g_i(x) = \text{score}_i(x) + \mathcal{N}(0, \sigma^2)
\label{eq11}
\end{equation}

where $\text{score}_i(x)$ denotes the learned relevance score, and $\mathcal{N}(0, \sigma^2)$ represents Gaussian noise injected to encourage exploration and prevent expert collapse. These scores are normalized using softmax to produce the routing probabilities $p_i(x)$, and only the top-$k$ experts are activated based on the highest probabilities.

To promote balanced expert utilization, an auxiliary load-balancing loss is introduced to penalize skewed routing distributions:
\begin{equation}
L_{\text{load}} = \sum_{i=1}^{N} \left(p_i(x) - \frac{1}{N}\right)^2
\label{eq12}
\end{equation}

where $N$ is the total number of experts. This loss penalizes skewed routing distributions (e.g., when 80\% of inputs are assigned to a single expert), promoting diversity and robustness in expert activation.

\begin{figure*}[htbp]
    \centering
    \includegraphics[height=0.45\textwidth]{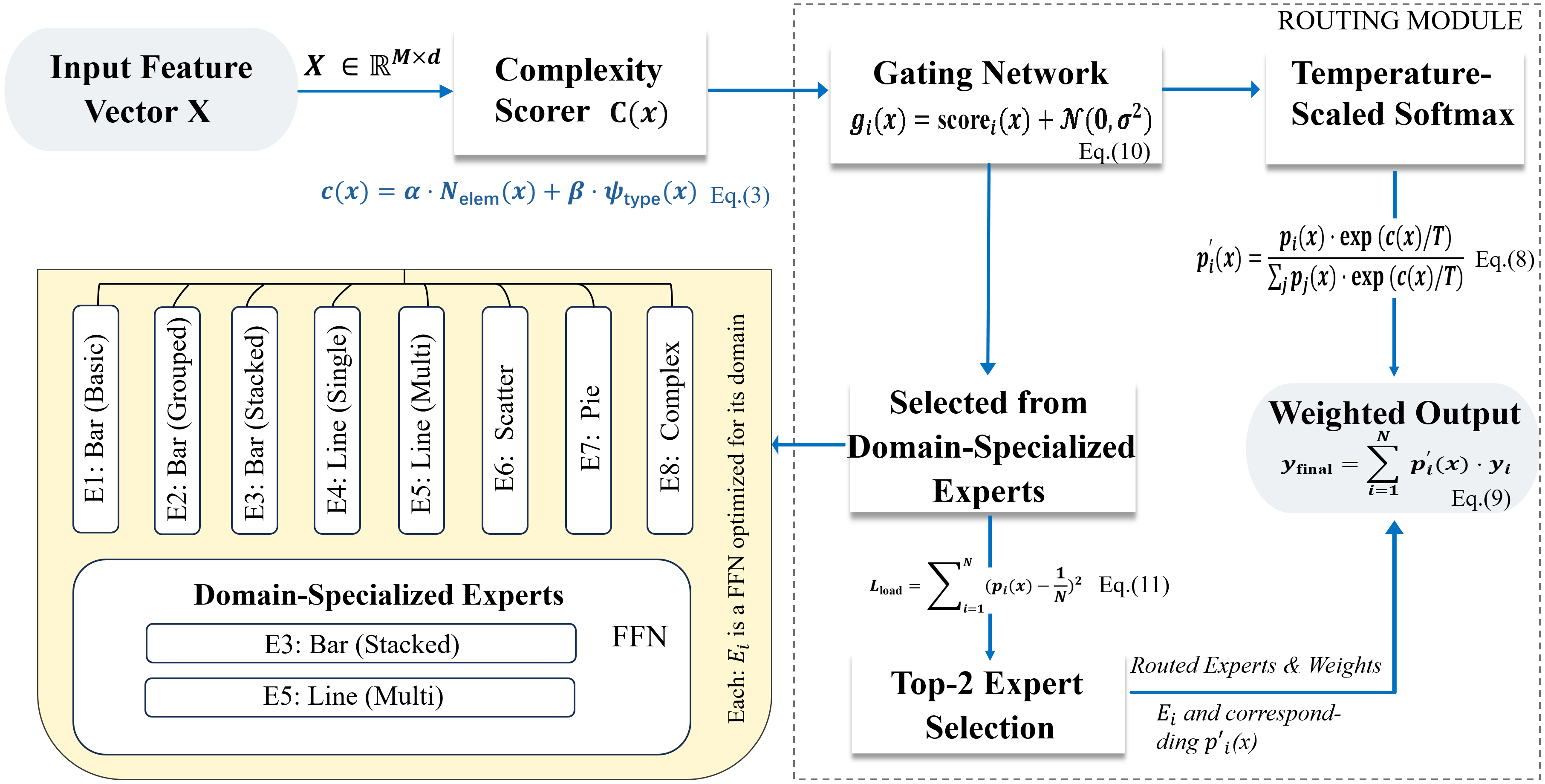}
    \caption{Expert Routing Mechanism in C2C-MoLA}
    \label{fig:2}
\end{figure*}

As shown in Fig.~\ref{fig:2}, the workflow of MoE's two components is divided into four stages, illustrating the routing process from input to expert selection.
First, the input feature vector $x \in \mathbb{R}^{M \times d}$ is scored via the complexity metric $c(x)$ (Eq.~\ref{eq:complexity}).
Second, a \textit{gating network} computes expert scores with Gaussian noise (Eq.~\ref{eq11}).
Third, these scores are normalized using \textit{temperature-scaled softmax} (Eq.~\ref{eq:complexity_softmax}) to obtain routing probabilities $p'_i(x)$, from which the top-2 domain experts are selected. 
Finally, expert outputs are fused via \textit{weighted sum} (Eq.~\ref{eq:expert_output}), and the regularization term  (Eq.~\ref{eq12}) promotes balanced expert usage.

\subsection{ LoRA-Based Parameter-Efficient Adaptation}
To reduce computational and memory overhead while preserving model performance, we employ a Low-Rank Adaptation (LoRA) strategy to fine-tune the MoE-enhanced chart-to-code model. As illustrated in Fig.~\ref{fig:1}, LoRA introduces a trainable low-rank update $\Delta W = AB$ to selected weight matrices $W$, such that:
\begin{equation}
W = W_0 + \Delta W = W_0 + AB
\label{eq13}
\end{equation}

where $A \in \mathbb{R}^{d \times r}$, $B \in \mathbb{R}^{r \times d}$, and $r \ll d$. In the implementation, LoRA is applied selectively to the attention projection matrices including  $W_q$, $W_k$, and $W_v$, while the feed-forward networks are kept frozen to minimize parameter overhead.

For stable optimization, the matrices are initialized with $A \sim \mathcal{U}(-0.01, 0.01)$ and $B = 0$, ensuring that the initial update $\Delta W$ is zero and does not perturb the pretrained weight $W_0$. The overall training objective incorporates a Frobenius-norm regularization to encourage low-rank expressiveness and prevent overfitting:
\begin{equation}
\min_{r} \mathcal{L}(r) = \mathcal{L}_{\text{task}} + \lambda \left(\|A\|_F^2 + \|B\|_F^2 \right)
\label{eq14}
\end{equation}

where $\lambda$ regulates the strength of regularization. To further stabilize optimization, gradient clipping, learning rate scheduling, and rank tuning are applied. Model parameters $\theta$ = \{A, B\} are updated using standard gradient descent:
\begin{equation}
\theta \leftarrow \theta - \eta \cdot \nabla_\theta \mathcal{L}
\label{eq15}
\end{equation}

This strategy enables efficient task-specific adaptation of large pre-trained models, achieving high performance with significantly reduced training cost.

\subsection{ Training Strategy}
To support stable and scalable optimization of the modular architecture, we design a customized training optimization strategy that facilitates stable expert routing and scalable training under memory constraints. This strategy plays a key role in enabling C2C-MoLA to realize its design objectives by aligning routing behavior with learning objectives and minimizing resource overhead. The pipeline consists of three integrated components: multi-task loss composition, optimization stabilization, and distributed memory reduction.

\subsubsection{Multi-Task Loss Composition}
The total loss integrates a primary code generation loss and two auxiliary components: routing loss and expert utilization regularization. Formally,
\begin{equation}
\mathcal{L}_{\text{total}} = \mathcal{L}_{\text{primary}} + \lambda_2 \mathcal{L}_{\text{router}} + \lambda_3 \mathcal{L}_{\text{util}},
\label{eq16}
\end{equation}

The primary loss is defined as $\mathcal{L}_{\text{primary}} = \mathcal{L}_{\text{syntax}} + \mathcal{L}_{\text{semantic}}$, ensuring both syntactic validity and visual fidelity. Specifically, $\mathcal{L}_{\text{syntax}}$ is the cross-entropy loss between the predicted and ground-truth code tokens, while 
$\mathcal{L}_{\text{semantic}}$ penalizes outputs whose rendered charts yield an IoU score below the threshold $\tau$.

The router loss $\mathcal{L}_{\text{router}}$ is defined as the Kullback–Leibler (KL) divergence between the predicted expert routing distribution and a uniform prior. This encourages the model to select experts meaningfully while maintaining diversity across samples.

The utilization regularization term encourages balanced usage across all experts. It is computed as:
\begin{equation}
\mathcal{L}_{\text{util}} = \lambda_1 \sum_{i=1}^n \left| u_i - \frac{1}{n} \right|^2,
\label{eq17}
\end{equation}

where $u_i$ denotes the usage rate of expert $i$, measured over every 1k training steps. We adopt $\lambda_2 = 0.7$ and $\lambda_3 = 0.3$ based on grid search optimization.

\subsubsection{Optimization Pipeline}
Training is performed using the AdamW optimizer with parameters $\beta_1 = 0.9$, $\beta_2 = 0.999$, and $\epsilon = 10^{-8}$, along with a weight decay coefficient of 0.01. A cosine annealing learning rate schedule is applied:
\begin{equation}
\eta_t = \eta_{\text{min}} + \frac{1}{2}(\eta_{\text{max}} - \eta_{\text{min}})\left(1 + \cos\left(\frac{t}{T_{\text{max}}} \pi\right)\right).
\label{eq18}
\end{equation}

where $\eta_{\max} = 10^{-4}$, $\eta_{\min} = 10^{-6}$, and $T_{\max} = 50{,}000$ steps.

To stabilize training and mitigate overfitting, dropout with a probability of $0.1$–$0.3$ is applied to feed-forward layers, and weight decay regularization is used across parameters $\theta$. Gradient clipping is employed to limit the magnitude of updates:
\begin{equation}
g_i = \frac{g_i}{\max\left(1, \frac{\|g_i\|}{c}\right)},
\label{eq19}
\end{equation}

Additionally, input augmentation is performed by applying random chart rotations ($\pm 15^\circ$) and color jittering ($\pm 10\%$) during training. Chart images are resized and normalized, while code sequences are tokenized into semantically meaningful units to improve visual-code alignment.

\subsubsection{Memory Optimization}
To support efficient training of the large-scale architecture (Qwen2.5-7B with MoE), memory optimization is achieved through three techniques. First, DeepSpeed ZeRO-3 is used to partition model parameters, gradients, and optimizer states across $K=8$ GPUs, enabling distributed training with minimal memory consumption per device. Formally, model weights and gradients are distributed as:
\begin{equation}
\theta = \bigcup_{k=1}^K \theta_k, \quad g = \frac{1}{K} \sum_k g_k,
\label{eq20}
\end{equation}

Second, gradient checkpointing is applied at block-level granularity (one checkpoint per four layers), reducing activation memory by approximately 47\%. This technique is selectively applied to core layers, excluding MoE experts to avoid unnecessary recomputation.

Third, mixed precision training is performed using BFloat16 for forward and backward passes, while maintaining optimizer states in Float32. This reduces memory consumption by nearly 47\% compared to full-precision training, without degrading convergence quality.

\section{Evaluation}
\label{sec:eva}
In this section, the proposed C2C-MoLA framework is evaluated to validate its practical effectiveness and address the following three research questions (RQs):

\begin{itemize}
\item \textbf{RQ1:} Does the proposed model achieve higher generation accuracy in chart-to-code tasks, particularly when handling complex chart types?
\item \textbf{RQ2:} Is the proposed model more efficient in terms of resource utilization, achieving lower memory consumption and faster convergence while maintaining high performance?
\item \textbf{RQ3:} Do the design and configuration of the MoE routing mechanism and LoRA parameterization support interpretability, efficiency, and optimal trade-offs in model performance?
\end{itemize}

To address RQ1, we compare the model’s execution success rate and AST-based semantic similarity across five representative chart categories (bar, line, pie, scatter, and complex charts), using both standard fine-tuning and LoRA-only approaches as baselines.

For RQ2, we analyze peak GPU memory usage, training convergence speed, and inference latency, verifying the efficiency advantages introduced by expert specialization and parameter-efficient adaptation.

To address RQ3, we conduct two complementary analyses to evaluate the design and configuration of the MoE routing mechanism and LoRA parameterization.
Sec.~IV-B.3 compares expert selection patterns with and without load-balancing regularization to assess whether the routing mechanism promotes balanced and potentially interpretable expert activation. Sec.~IV-B.4 presents ablation studies quantifying the performance impact and computational trade-offs across different MoE configurations (expert count, token capacity, routing strategy) and LoRA settings (rank, target modules, $\alpha$).

\subsection{Settings}
\subsubsection{Dataset}
We conduct experiments on Chart2Code-160k\cite{yang2024chartmimic}, a large-scale dataset specifically curated for chart-to-code generation research. It contains $160,000$ chart and code pairs, each consisting of a chart image and the corresponding Python visualization code. The dataset spans five representative chart types, covering a broad range of visual and structural complexity seen across chart designs.

To ensure robust evaluation, we adopt a stratified $70$/$15$/$15$ split for training, validation, and testing, preserving the chart-type distribution across subsets to reduce sampling bias, as shown in Tab.~\ref{tab:1}. 
Complex and rare layouts, including $32$ variants such as stacked area and dual-axis charts, are reserved for testing to assess generalization under challenging conditions.
\begin{table}[htbp]
\centering
    \caption{Distribution of Chart Types Across Dataset Splits} 
    \begin{tabular}{c|c|c|c|c}
    \toprule
        Chart Type & Training Set & Validation Set & Test Set & Complexity\\
    \midrule
        Bar Chart & 35{,}000 & 5{,}000 & 5{,}000 & 3.2 \\
        Line Chart & 30{,}000 & 5{,}000 & 5{,}000 & 4.1 \\
        Scatter Plot & 20{,}000 & 3{,}000 & 3{,}000 & 4.5 \\
        Pie Chart & 15{,}000 & 2{,}000 & 3{,}000 & 2.7\\
        Mixed Charts & 10{,}000 & 2{,}000 & 2{,}000 & Adaptive \\
        \textbf{Total} & \textbf{112{,}000} & \textbf{24{,}000} & \textbf{24{,}000}\\
    \bottomrule
    \end{tabular}
\caption*{\footnotesize * \textit{Complexity} is computed via Eq.\ref{eq:complexity} using element count and chart-type weight.}
    \label{tab:1}
\end{table}

Chart images are standardized to a resolution of $512$×$512$, and the corresponding code sequences have an average length of $125$ tokens.
To improve generalization and mitigate overfitting to common patterns, we apply data augmentation at both image and code levels.
For chart images, we introduce geometric transformations such as rotation, scaling, translation, and visual noise to simulate real-world variations. 
For code, we apply functionality-preserving modifications to variable names, function structures, and comments to increase lexical and structural diversity.

\subsubsection{Baselines and Variants}
To evaluate the effectiveness of C2C-MoLA, three representative baselines are compared under consistent training settings, as summarized in Tab.~\ref{tab:2}. All models are trained using the same dataset splits and preprocessing pipeline. Where applicable, learning rates, batch sizes, and stopping criteria are aligned to ensure fair comparison.

\textbf{Standard Fine-Tuning}: The full model is trained via backpropagation without applying MoE or LoRA. This serves as a conventional benchmark for isolating the contribution of expert routing and parameter-efficient adaptation.

\textbf{LoRA-Only Adaptation}: Low-rank adaptation is applied to attention projections (i.e., $W_q$, $W_k$, $W_v$) without using the MoE module. This setting isolates the effect of LoRA in the absence of expert routing.

\textbf{Existing Chart2Code Models}: ChartLLaMA and ChartCoder are included as external baselines. These transformer-based encoder-decoder models are trained on the same dataset using their original procedures. 
However, due to incompatible optimization settings and limited reproducibility, they are not included in the quantitative comparison.
\begin{table*}[htbp]
\centering
    \caption{Baseline Model Comparison} 
    \begin{tabular}{c|c|c|c|c}
    \toprule
        \multirow{2}{*}{Model} & Memory & Computational & Chart Type & Notable\\
             & Efficiency & Overhead & Handling & Features\\
     \midrule
        Standard Fine-Tuning & Low & High & Limited to simpler charts & Full parameter updates \\
        LoRA-Only Approach & High & Moderate & Handles a variety of charts with efficiency & Low-rank adaptation \\
        Existing Chart2Code & Moderate & High & Varies with chart complexity & Transformer-based image-to-text models \\
    \bottomrule
    \end{tabular}
    \label{tab:2}
\end{table*}

\subsubsection{Implementation Details}
Experiments are conducted on NVIDIA A100 GPUs (40 GB each), connected via a high-speed InfiniBand network using the NCCL backend for dis-
tributed training. To enable scalable and memory-efficient training, we adopt DeepSpeed ZeRO-3 for parameter and optimizer state partitioning, along with gradient checkpointing to reduce activation memory. Mixed-precision training with BFloat16 is applied throughout to further reduce memory usage without compromising accuracy.

Key hyperparameters, including learning rate, optimizer, dropout, expert configuration, and routing thresholds, are summarized in Tab.~\ref{tab:3}. These values were selected via grid search to optimize the interaction between LoRA adaptation and MoE routing, ensuring stable convergence and efficient expert utilization. The final MoE and LoRA settings, determined based on ablation results (Sec.~IV-B.4), include 8 experts with a token capacity of 32 and top-2 routing for MoE, and a LoRA rank of 8 applied to both attention and MLP modules with a scaling factor $\alpha=16$.
\renewcommand{\arraystretch}{1.15}
\begin{table}[htbp]
\centering
\caption{Training Hyperparameter Settings}
\begin{tabular}{l|l}
\toprule
\textbf{Configuration} & \textbf{Details} \\
\midrule
\multicolumn{2}{l}{\textit{General Training Settings}} \\
\midrule
Batch Size & 4 per device \\
Learning Rate & 1e-4 (optimized for LoRA + MoE) \\
Optimizer & AdamW \\
Scheduler & Cosine annealing \\
Dropout Rate & 0.1 \\
Precision & BFloat16 (mixed precision) \\
Gradient Checkpointing & Enabled (for memory-time trade-off) \\
\midrule
\multicolumn{2}{l}{\textit{MoE Configuration}} \\
\midrule
Number of Experts & 8 \\
Expert Capacity & 32 tokens per expert \\
Routing Strategy & Top-2 gating \\
Routing Temperature $T$ & 1.0 (empirically tuned) \\
Routing Noise $\sigma$ & 0.01 \\
Utilization Reg. $\lambda_1$ & 0.5 \\
\midrule
\multicolumn{2}{l}{\textit{LoRA Configuration}} \\
\midrule
LoRA Rank & 8 \\
Target Modules & Attention + MLP \\
Scaling Factor $\alpha$ & 16 \\
\midrule
\multicolumn{2}{l}{\textit{Evaluation Threshold}} \\
\midrule
IoU Threshold $\tau$ & 0.85 (fidelity criterion) \\
\bottomrule
\end{tabular}
\label{tab:3}
\end{table}

\subsubsection{Evaluation Metrics}
To comprehensively assess model performance, we adopt metrics spanning three key dimensions: code quality, computational efficiency, and visual-semantic alignment, as summarized in Tab.~\ref{tab:4}.

\textbf{Code quality} reflects the syntactic correctness, functional validity, and structural fidelity of the generated code. Outputs are evaluated based on their conformity to Python grammar and their ability to render correctly \cite{evtikhiev2023out}. 
A generation is considered successful if its rendered chart matches the input with an IoU~$\geq$~0.85. Formally, the success rate is defined as:
\begin{equation}
\text{Success Rate} = \frac{1}{N} \sum_{i=1}^{N} \mathbf{1}\left\{ \text{IoU}(I_i, I_i') \geq \tau \right\}
\label{eq:success_rate}
\end{equation}
where \( \tau = 0.85 \) and $\mathbf{1}\{\cdot\}$ denotes the indicator function.

\textbf{Computational efficiency} covers both training and inference resource usage, including peak GPU memory, convergence time (in epochs), and inference latency per chart. Special attention is given to the overhead introduced by the MoE routing mechanism.

\textbf{Visual-semantic alignment} is implicitly assessed through the IoU-based success rate, which reflects how well the generated code preserves the structural and visual characteristics of the input chart\cite{han2023chartllama}.
\renewcommand{\arraystretch}{1.188}
\begin{table}[htbp]
\centering
\caption{Evaluation Metrics Overview}
\begin{tabular}{p{1.4cm}|p{2.3cm}|p{3.7cm}}
\toprule
Category & Metric& Definition \\
\midrule
\multirow{1}{*}{\raisebox{-2.6\height}{Code Quality}}
& \multirow{1}{*}{\raisebox{-2.6\height}{Success Rate}} & Ratio of code whose rendered charts match input (IoU~$\geq$~0.85). Ref
lects syntax and structural fidelity.\\
\midrule
\multirow{4}{*}{Efficiency}
& Peak Memory & Max GPU and gradient cost. \\
& Training Time & Epochs to convergence.\\
& Inference Latency & Per-chart generation time. \\
& Routing Overhead & Extra latency from MoE routing. \\
\midrule
\multirow{1}{*}{\raisebox{-2\height}{Visual Match}}
& \multirow{1}{*}{\raisebox{-2\height}{IoU Match}} & Measures structural alignment between input and rendered charts via IoU threshold. \\
\bottomrule
\end{tabular}
\label{tab:4}
\end{table}

\subsection{Results}
\subsubsection{Overall Performance}
We evaluate the overall performance of the C2C-MoLA framework across five chart types: bar, line, pie, scatter, and complex charts. Evaluation is based on success rate (rendered IoU~$\geq$~0.85) and visual alignment with the input chart, reflecting both syntactic correctness and semantic fidelity.

As shown in Tab.~\ref{mix1}, C2C-MoLA consistently outperforms both standard fine-tuning and LoRA-only baselines across all chart types. 
Notably, it achieves success rates of 92\% on bar charts and 91\% on scatter plots, outperforming the baselines by 7–10\%. 
Complex charts, which involve multi-series or overlapping elements, benefit most from adaptive expert routing, achieving a 17\% improvement over standard fine-tuning.
\begin{table}[htbp]
\centering
\caption{Comparison of Accuracy and Resource Efficiency}
\label{tab:chart_comparison_simple}
\begin{tabular}{c|c|c|c|c}
\toprule
Metric & MoE & Standard & LoRA &  $\Delta$ vs.Standard\\

\midrule
\multicolumn{5}{c}{\textit{Success rate(\%)}} \\
Bar Chart & \textbf{92} & 85 & 88 & +7 \\
Line Chart & \textbf{90} & 80 & 85 & +10 \\
Pie Chart & \textbf{89} & 75 & 80 & +14 \\
Scatter Plot & \textbf{91} & 80 & 84 & +11 \\
Complex Charts & \textbf{87} & 70 & 76 & +17 \\
\midrule

\multicolumn{5}{c}{\textit{Training}} \\
Peak Memory (GB) & \textbf{12.3} & 15.0 & 13.5 & 4.84 \\
Epochs & \textbf{4} & 5 & \textbf{4} & 1 \\
Convergence Time (h) & \textbf{4} & 5 & 4.5 & - \\
\midrule

\multicolumn{5}{c}{\textit{Inference}} \\
Latency (s) & \textbf{0.45} & 0.40 & 0.42 & +0.05s \\
Route Overhead (s) & \textbf{+0.05} & 0 & +0.02 & - \\
\bottomrule

\end{tabular}
\label{mix1}
\end{table}

To assess statistical significance, we conduct paired t-tests comparing the C2C-MoLA model with each baseline. The results confirm that the accuracy improvements are statistically significant (p $<$ $0.05$) across all chart types, indicating that the performance gains stem from architectural innovations rather than random variation.

\subsubsection{Efficiency Metrics}
We evaluate the efficiency of C2C-MoLA against standard fine-tuning and LoRA-only baselines across three critical dimensions: peak GPU memory consumption, convergence speed, and inference latency including routing overhead. A summary of results is shown in Tab.~\ref{mix1}.

During training, the C2C-MoLA model reduces peak GPU memory usage by approximately 18\% (from $15.0$GB to $12.3$GB) compared to standard fine-tuning, primarily due to selective expert activation and mixed-precision computation (BFloat$16$). 
The model also reaches convergence 20\% faster ($4$ vs. $5$ epochs), likely benefiting from LoRA’s low-rank updates, which support more efficient gradient propagation.

During inference, the MoE model incurs 5\% routing overhead, increasing per-chart latency by $0.05$ seconds compared to the baseline. This slight increase results from the expert selection mechanism introduced by the MoE architecture.

\begin{figure}[htbp]
    \centering
    \includegraphics[width=0.48\textwidth]{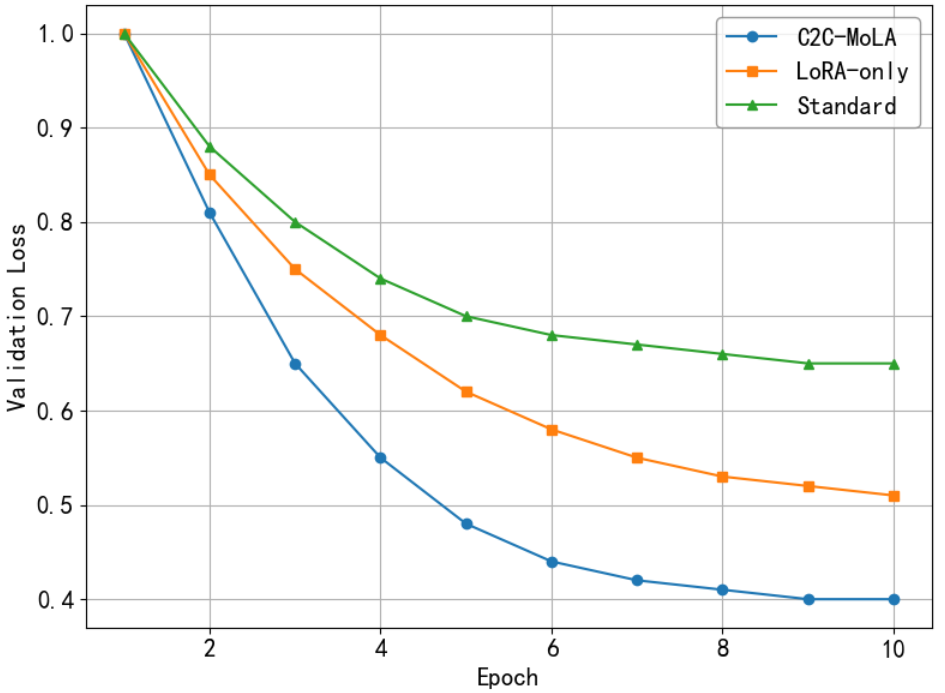}
    \caption{Training Convergence Curves}
    \label{fig:convergence_curve}
\end{figure}

Overall, these results show that although the MoE design introduces a slight increase in inference time, it yields clear gains in training efficiency by reducing memory usage and accelerating convergence, making it well suited for deployment under constrained computational resources. 
As shown in Fig.~\ref{fig:convergence_curve}, this C2C-MoLA model not only achieves faster convergence, but also demonstrates more stable training compared to LoRA-only and standard baselines.

\subsubsection{Expert Utilization}
\begin{figure}[htbp]
    \centering
    \includegraphics[width=0.48\textwidth,height=0.22\textheight]{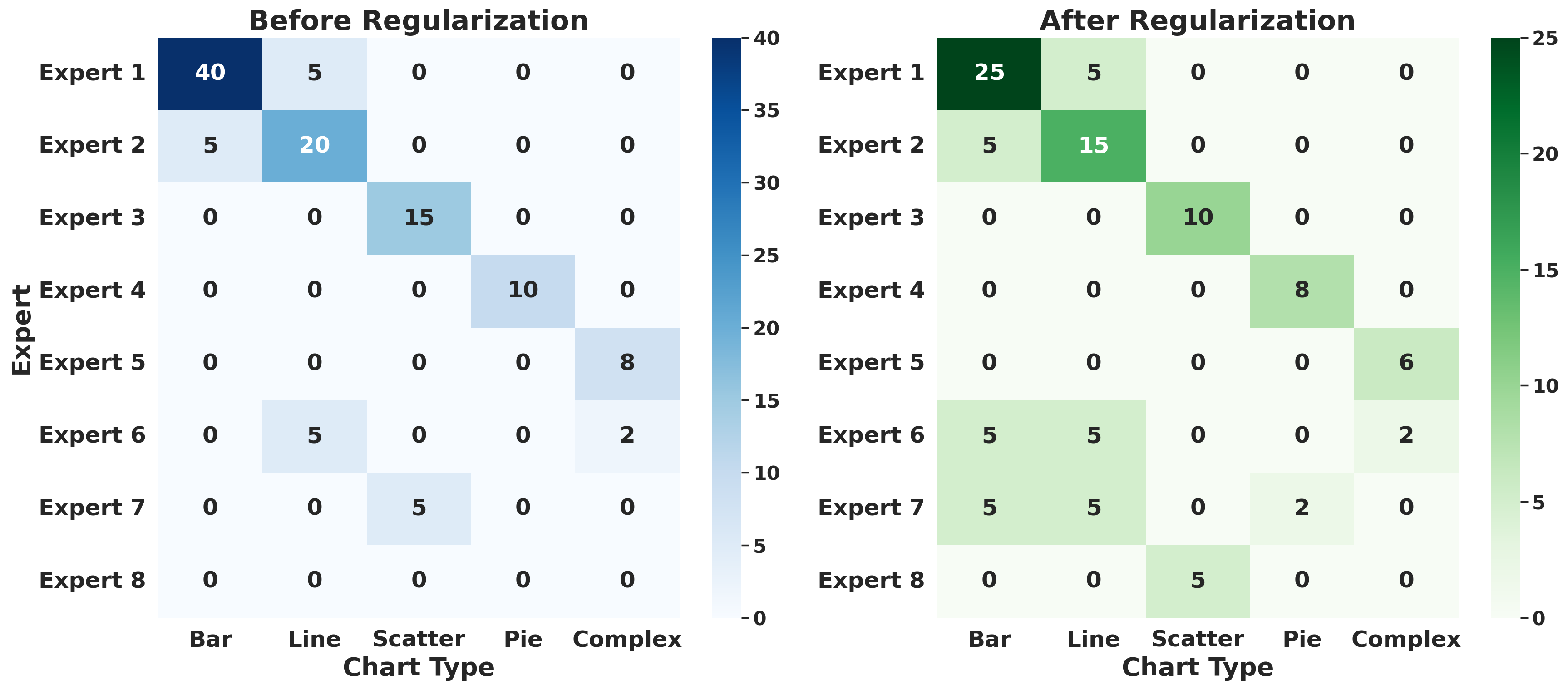}
    \caption{Expert selection before vs. after regularization}
    \label{fig:expert_heatmap}
\end{figure}

To assess specialization and load balancing in the MoE architecture, we analyze expert selection patterns across chart types, as visualized in Fig.~\ref{fig:expert_heatmap}.

Before regularization, expert selection was highly skewed. For example, expert $1$ handled 40\% of bar chart inputs, more than double that of any other expert.
Expert $3$ and Expert $4$ accounted for 15\% of scatter plots and 10\% of pie charts, respectively. 
Several experts, including Expert 5 through Expert 8, exhibited minimal or no activation, indicating significant load imbalance.

After applying auxiliary load balancing regularization, expert usage becomes more evenly distributed. Expert1’s load for bar charts drops to 25\%, while Expert3’s scatter plot assignment decreases to 10\%. 
Previously underutilized experts, such as Expert $6$ and Expert $7$, were activated across multiple chart types, each contributing between 2\% and 5\%. Expert $8$, previously unused, began handling 5\% of scatter plots.

This redistribution reduces the dominance of individual experts and increases overall participation. The selection matrix shows broader engagement across all eight experts and five chart types, reflecting improved task coverage and more balanced routing, which supports more stable and scalable MoE training.

\subsubsection{Ablation Studies}
We conduct comprehensive ablations to validate the design choices of MoE and LoRA described in Sec.~III.
Results in Tab.~\ref{mix2} identify optimal configurations that balance accuracy and resource efficiency.

\begin{table*}[htbp]
\centering
\caption{Ablation Results for MoE and LoRA}
\label{tab:comp_config}

\begin{tabular}{c|ccc|ccc|cc|ccc|ccc|ccc}
\toprule
\multirow{3}{*}{Parameters} & \multicolumn{8}{c|}{\textbf{Ours MoE}} & \multicolumn{9}{c}{\textbf{LoRA}} \\
\cmidrule(lr){2-9} \cmidrule(lr){10-18}

& \multicolumn{3}{c|}{Experts} & \multicolumn{3}{c|}{Capacity} & \multicolumn{2}{c|}{Routing}
& \multicolumn{3}{c|}{Rank} & \multicolumn{3}{c|}{Alpha} & \multicolumn{3}{c}{Target Module} \\

& 4 & \cellcolor{gray!20}8 & 12 & 16 & \cellcolor{gray!20}32 & 64 & Top-k & \cellcolor{gray!20}Prob.
& 4 & \cellcolor{gray!20}8 & 16 & 16 & \cellcolor{gray!20}32 & 64 & Attn & Out & \cellcolor{gray!20}

Attn+Out \\
\midrule
Accuracy (\%) & 45 & 57 & 59 & 43 & 57 & 59 & 52 & 57 & 48 & 58 & 60 & 55 & 58 & 60 & 52 & 58 & 60\\
Peak Memory (GB) & - & - & - & 15 & 22 & 30 & - & - & 8 & 12 & 20 & 10 & 12 & 16 & - & - & - \\
Training Time (h) & - & - & - & 10 & 25 & 20 & - & - & 10 & 15 & 25 & 12 & 15 & 18 & - & - & - \\
Overhead (\%) & 25 & 35 & 45 & - & - & - & 20 & 25 & - & - & - & - & - & - & - & - & - \\
Util. Balance (\%) & 70 & 65 & 60 & - & - & - & 75 & 65 & - & - & - & - & - & - & - & - & -  \\
Overfit (\%) & - & - & - & - & - & - & - & - & - & - & - & - & - & - & 5 & 8 & 20\\
Stability (\%) & - & - & - & - & - & - & - & - & - & - & - & - & - & - & 90 & 95 & 85 \\
\bottomrule
\end{tabular}

\vspace{1mm}

\begin{minipage}{0.97\textwidth}
\footnotesize
\textbf{*} 
Routing: MoE strategy (Top-k = top-k experts; Prob. = probabilistic routing).
Target Module: LoRA injection scope (Attn = attention; Out = output; Attn+Out = both).
Overhead: MoE routing overhead (\%).
Util.Balance: expert utilization balance (\%).
“-”: not measured or not applicable.
\end{minipage}
\label{mix2}
\end{table*}

\paragraph{MoE Configuration Analysis}
Three key MoE parameters are evaluated: expert count, token capacity, and routing strategy. Increasing experts from $4$ to $8$ improves accuracy from 45\% to 57\% with moderate memory overhead (35\%), Using $12$ experts yields only a marginal gain (59\%) but increases overhead to 45\%. 
A capacity of $32$ tokens per expert achieves optimal accuracy (57\%), outperforming $16$ tokens (43\%) with a 22\% memory increase. Probabilistic routing delivers better accuracy (57\% vs. 52\%) than top-k, albeit with slightly higher overhead (25\% vs. 20\%).

\paragraph{LoRA Configuration Analysis}
We analyze the impact of LoRA rank, target modules, and alpha. Rank-8 achieves a strong balance between performance and training cost, improving accuracy to 58\%, compared to 48\% for rank-4. 
Rank-16 offers a marginal gain (60\%) but incurs significantly higher memory and training time.
Targeting both attention and output modules yields the highest accuracy (60\%), surpassing attention-only (52\%) and output-only (58\%) settings. 
For alpha, a value of $32$ achieves the best trade-off, reaching 58\% accuracy with moderate overfitting risk (8\%) and high training stability (95\%). Higher alpha values (e.g., $64$) introduce greater overfitting and instability.

Ablation results show diminishing returns beyond $8$ experts or LoRA rank-8, accompanied by increased memory usage, overhead, or overfitting. The configuration with $8$ experts, 32-token capacity, and probabilistic routing, combined with LoRA rank-8 (targeting both attention and output, alpha = $32$), yields competitive accuracy with manageable resource requirements. This setup provides a balanced option for practical use.

\subsection{Discussions}
\subsubsection{ Generation Accuracy and Adaptability}
The proposed C2C-MoLA model demonstrates consistent improvements in generation accuracy across diverse chart types, particularly under high visual or structural complexity.
The dynamic routing mechanism plays a critical role in enabling specialized expert activation, enabling the model to better capture chart semantics and graphical nuances.
Compared to both standard fine-tuning and LoRA-only baselines, the model achieves statistically significant performance gains, confirmed through paired t-tests. In bar charts, scatter plots, and complex visualizations, accuracy improvements range from 7\% to 17\%, accompanied by higher execution success rates and improved AST similarity. 
These findings highlight the model’s ability to generalize across diverse visual patterns while preserving semantic fidelity in code generation.

\subsubsection{Resource Efficiency and Scalability}
The combination of MoE and LoRA leads to notable resource savings during training. Through selective expert activation and mixed-precision computation, peak GPU memory usage is reduced by approximately 18\% ($12.3$ GB vs. $15.0$ GB). Training convergence is also accelerated, requiring 20\% fewer epochs compared to standard fine-tuning. 
While inference latency increases slightly by $0.05$ seconds per chart (5\% overhead), the added cost is largely offset by faster training and improved expert utilization. 
These results demonstrate favorable trade-offs between performance and efficiency, supporting deployment in resource-constrained or latency-sensitive environments.

\subsubsection{Component Effectiveness and Design Trade-offs}
Component-level evaluations validate the effectiveness of the MoE and LoRA design choices.
Expert selection analysis confirms that routing behavior varies with chart type and complexity, while auxiliary load balancing promotes more uniform expert utilization.
Ablation studies reveal that scaling beyond 8 experts or LoRA rank $8$ yields diminishing accuracy improvements (less than 2\%) at the cost of significantly increased resource usage. 
A token capacity of $32$ and probabilistic routing provide the best trade-off between performance and overhead.
For LoRA, adapting both attention and output layers with rank-8 and $\alpha=32$ provides a strong balance between accuracy and overfitting risk.

Despite the additional inference overhead introduced by routing, latency remains acceptable for practical use (0.45s per chart), supporting the feasibility of real-world deployment.
Overall, the combined MoE-LoRA architecture delivers interpretable, efficient, and accurate code generation for diverse chart inputs.

\subsection{Validity Analysis}
\textbf{Internal validity.} Our experiments are conducted on the Chart2Code160k dataset, which is synthetically generated. 
Although the dataset covers a broad range of chart types and structural variations, it may not fully capture the complexity and noise present in real-world settings, such as ambiguous annotations, irregular layouts, or graphical artifacts. 
While we employ data augmentation and stratified sampling to mitigate overfitting and enhance robustness, generalization to non-synthetic or visually noisy charts remains an open challenge.

\textbf{External validity.} The proposed model is primarily evaluated on charts rendered from well-structured Python code using standard visualization libraries. Its performance on out-of-distribution charts, including hand-drawn sketches, scanned documents, or infographic-style visualizations, has not yet been assessed. Moreover, the model's reliance on clean and structured layouts may limit its applicability in domains with unstructured or heterogeneous visual inputs.

\textbf{Implementation validity.} To ensure reproducibility and fair comparison, all models are trained under consistent settings, and hyperparameters are selected with held-out validation data. Baselines are re-run using our unified codebase.

\section{Conclusion}
This study presents C2C-MoLA, a MoE-enhanced chart-to-code generation framework that combines adaptive expert routing with parameter-efficient LoRA fine-tuning.
Evaluated on the Chart2Code-160k benchmark, the proposed model achieves 7\%--17\% higher generation accuracy than standard baselines,with particularly strong gains on structurally complex charts.
It also reduces peak GPU memory usage by 18\% and accelerates convergence by approximately 20\% compared to full fine-tuning.
Ablation results confirm the effectiveness of key configurations: $8$ experts with a 32-token capacity and probabilistic routing, combined with a LoRA setup using rank-8 and $\alpha=32$ applied to both attention and output layers.
These findings demonstrate that the proposed architecture achieves a practical balance between accuracy, efficiency, and extensibility, enabling scalable deployment.

Future work will proceed in three directions. 
First, we plan to explore expert-centric pretraining, where individual experts are initialized on domain-specific chart types (e.g., a scatter-plot expert pretrained on correlation-heavy data), to reduce routing ambiguity and improve specialization stability.
Second, to enhance robustness, we aim to extend the framework to handle noisy or real-world charts, including hand-drawn and scanned inputs, through sketch-augmented training and distortion-tolerant encoders. 
Third, we will investigate human-in-the-loop routing, integrating reinforcement learning and user feedback (e.g., preference rankings) to improve interpretability and adaptability in interactive or high-stakes settings. These directions aim to improve the generalizability, robustness, and real-world deployability of visual-to-code generation systems.

\bibliographystyle{IEEEtran}
\bibliography{main}

\end{document}